\documentclass[twocolumn,showpacs,preprintnumbers,amsmath,amssymb]{revtex4}

\usepackage{amsmath}
\usepackage{amssymb}
\usepackage{graphicx}% Include figure files

\begin{document}

\title{Phase-coding quantum-key-distribution system based on Sagnac–Mach-Zehnder interferometers}% Force line breaks with \\
\author{Xiao-Tian~Song$^*$, Dong~Wang$^*$, Xiao-Ming~Lu\footnote{The authors Xiao-Tian Song, Dong Wang and Xiao-Ming Lu contributed equally to this work.}, Da-Jun~Huang, Di~Jiang, Li-Xian~Li, Xi~Fang, Yi-Bo~Zhao\footnote{Corresponding author: zhaoyibo@mail.ustc.edu.cn}, and Liang-Jiang~Zhou}
\address{National Key Laboratory of Microwave Imaging Technology, Institute of Electronics, Chinese Academy of Sciences, Beijing 100190, China \\}

\date{\today}% It is always \today, today,

\begin{abstract}
  Stability and robustness are important criteria to evaluate the performance of a quantum key distribution (QKD) system in real-life applications. However, the inherent birefringence effect of the fiber channel and disturbance caused by the variation of the deployment environment of the QKD system tremendously decrease its performance. To eliminate this adverse impact, we propose a polarization-insensitive phase-coding QKD system based on Sagnac-Mach-Zehnder interferometers. Verified by theoretical analysis and experimental tests, this QKD system is robust against channel polarization disturbance. The robustness and long-term stability of the QKD system is confirmed with a 10-days continuous operation over a 12.6-dB channel, which consists of a 50-km fiber spool and a polarization scrambler (2 rad/s). As a result, an average quantum bit error rate of 0.958$\%$ and a sustained secure key rate of 3.68 kbps are obtained. Moreover, the secure key rate of the QKD system for a typical channel loss of 10 dB reaches 6.89 kbps, and the achievable maximum transmission distance exceeds 125 km.
\end{abstract}
\maketitle

%\section{Introduction}
\section{Introduction}
Quantum key distribution (QKD) \cite{BB84} can provide information-theoretic secure communication for two remote users. Over the past three decades, many QKD protocols have been developed and implemented in laboratory demonstrations \cite{decoy1,MDI,RRDPS,TF,exp1,exp2,exp3,exp4}. Up to date, the transmission distance of QKD has been extended to 421 km in optical fiber \cite{exp5} and the key generation rate has been raised to tens of megabits per second \cite{exp6}. Moreover, QKD has already come out from laboratories to field deployments over telecom networks \cite{network1,network2,network3,network4,network5,network6}, aiming for compatibility with installed fiber telecommunication networks.

In real-world deployment of QKD systems, the field environments of installed fibers used as quantum channels are usually complex and erratic \cite{DYY}. When photons are transmitted through a fiber channel, the intrinsic birefringence of single-mode fiber (SMF) causes the photons’ polarization to undergo a transformation, which varies with environmental fluctuations, such as temperature and other environmental influences. Therefore, photons’ polarization states become unpredictable when they arrive at the receiver, leading to a deterioration of the performance of polarization-sensitive QKD systems. This is clearly true for polarization-coding systems \cite{pol1,pol2,pol3}, but it is also a concern for phase-coding systems \cite{GYS,MZ} since the interference fringe visibility depends on the polarization alignment.

As for the phase-coding QKD system, the polarization-sensitivity feature of the original asymmetric Mach-Zehnder interferometer (AMZI) leads to its impracticality in field tests\cite{MZ}. Fortunately, efforts by researchers have been devoted to developing new methods to cope with the polarization effect of the fiber channel. Among these countermeasures, active polarization compensating components are often adopted in some polarization-dependent phase-coding systems \cite{YZL}, which are complex and time-consuming. In some other phase-coding systems, birefringence variations in fiber channels can be automatically compensated for, the schemes of which are carefully designed, such as the ``plug-and-play'' scheme \cite{Plug1} and Faraday-Michelson (FM) scheme \cite{MXF}. These schemes show excellent long-term stability, but suffer from speed limitation due to the round-trip structure and to potential Trojan-horse attack on the former one \cite{Trojan}. Another solution is to add a depolarizer at Alice's site, randomizing the polarization of the photons entering the fiber channel, and then a polarization beam splitter (PBS) is used by Bob to randomly route the depolarized photons into one of the two fixed AMZI \cite{Depol}. Nevertheless, it requires two interferometers and four single-photon detectors in the receiver, increasing the complexity of the system. Furthermore, the passive basis choice made by Bob's PBS may open a loophole for the intercept-resend attack \cite{IR}.

Recently, the Sagnac configuration has attracted lots of attention in this field owing to its intrinsically stable nature. Sagnac-based devices have been proposed for intensity modulation \cite{IM}, polarization modulation \cite{pol4,pol5}, as well as time-bin phase modulation \cite{WS}. In Ref. \cite{WS}, Wang \emph{et al.} improved the structure of a FM interferometer (FMI) by replacing one arm of the FMI with a Sagnac phase modulator, and they proposed a QKD scheme based on Faraday-Sagnac-Michelson interferometers (FSMIs), which is intrinsically stable against channel disturbance and has high-speed support.

Unlike Wang's proposal \cite{WS}, here in this paper, we put forward an alternative solution, by combining the Sagnac configuration with the conventional AMZI, to circumvent the effect of birefringent and disturbed fiber channel and other aforementioned problems. In our scheme, the sender remains the same as the conventional AMZI based QKD scheme, while the receiver is based on the SMZI. With the proposed scheme, a QKD system is built and the performance is demonstrated for long term stability.

The paper is organized as follows. In Sec. II, the scheme of the SMZI-based QKD system is introduced and its stability against polarization disturbance is analyzed. Sec. III describes the experimental setting. Sec. IV gives the experimental results and discussion. The article ends with some concluding remarks. The gains and error rates model in the simulation can be found in the Appendix.

\section{Scheme of the SMZI-based QKD}

\begin{figure*}[!ht]
\centering
\includegraphics[width=0.8\linewidth]{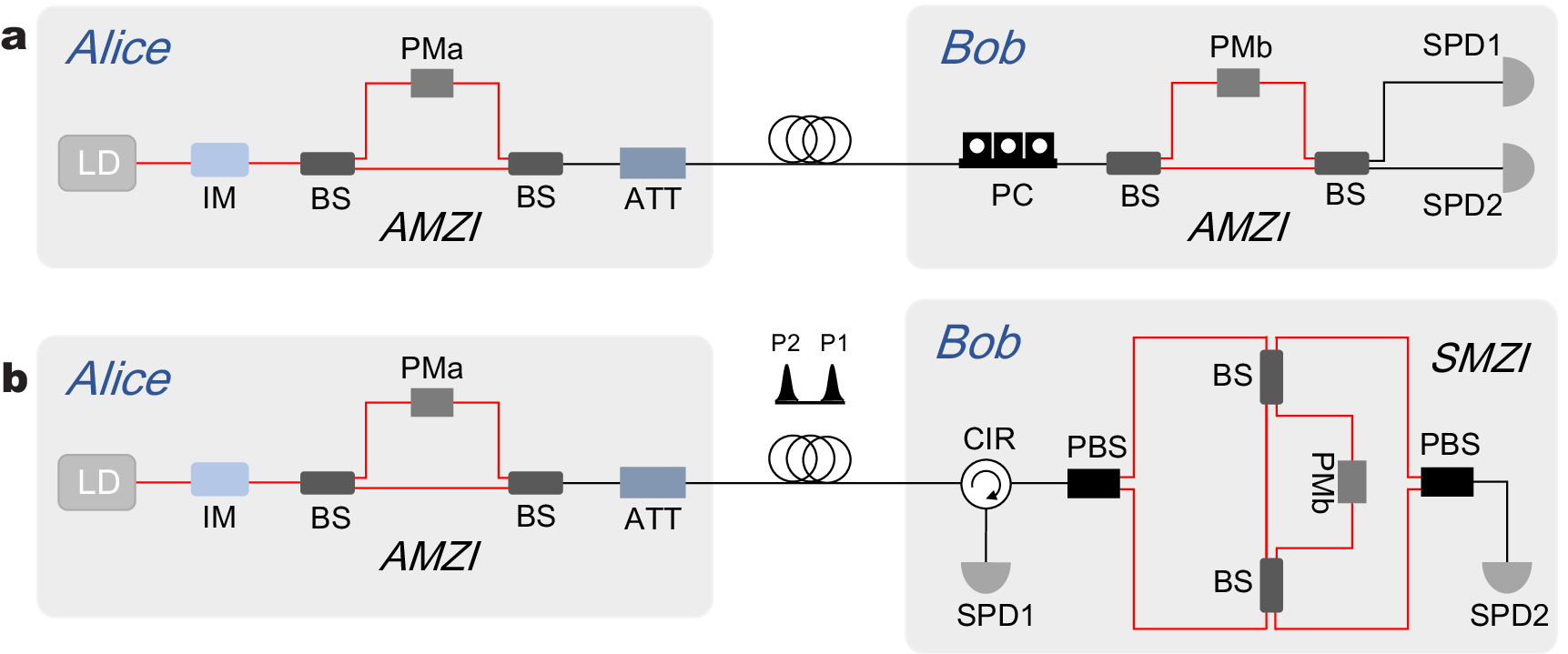}
\caption{a. Schematic of the AMZI-based QKD system; b. Schematic of the SMZI-based QKD system. LD, laser diode; IM, intensity modulator; PM, phase modulator; ATT, attenuator; PC, polarization controller; CIR, circulator; PBS, polarization beam splitter; SPD, single-photon detector. The red lines represent PMFs, and the black lines represent SMFs.}
\label{Fig:scheme}
\end{figure*}

A typical phase-coding BB84 QKD system based on the conventional AMZI is shown in Fig. \ref{Fig:scheme} a. Alice generates phase-randomized optical pulses with a laser diode (LD) and modulates the intensity of them with an intensity modulator (IM) to generate signal or decoy states required for the decoy-state method \cite{3decoy}. Then the pulses are encoded with four BB84 states by the phase modulator (PM) in her AMZI and attenuated to the single-photon level before entering a single-mode fiber channel. The pulses propagating along the short arm and the long arm of the AMZI are denoted by $P1$ and $P2$, respectively. The identical polarization of them is guaranteed by using polarization-maintaining fibers (PMFs) from the LD to the AMZI, as well as inside the AMZI. After transmission, photons received by Bob are decoded by an AMZI identical to the one in Alice and eventually detected by two single-photon detectors (SPDs). In order to obtain a high-interference fringe visibility, Bob should make use of a polarization controller (PC) to actively compensate the polarization drift of the received photons that are channel disturbed. While in the proposed SMZI-based QKD scheme, as demonstrated in Fig. \ref{Fig:scheme} b, the sender is not changed, and the receiver is remolded to be polarization-insensitive, thus removing the requirement of active polarization calibration.

\begin{figure}[!ht]
\centering
\includegraphics[width=1\columnwidth]{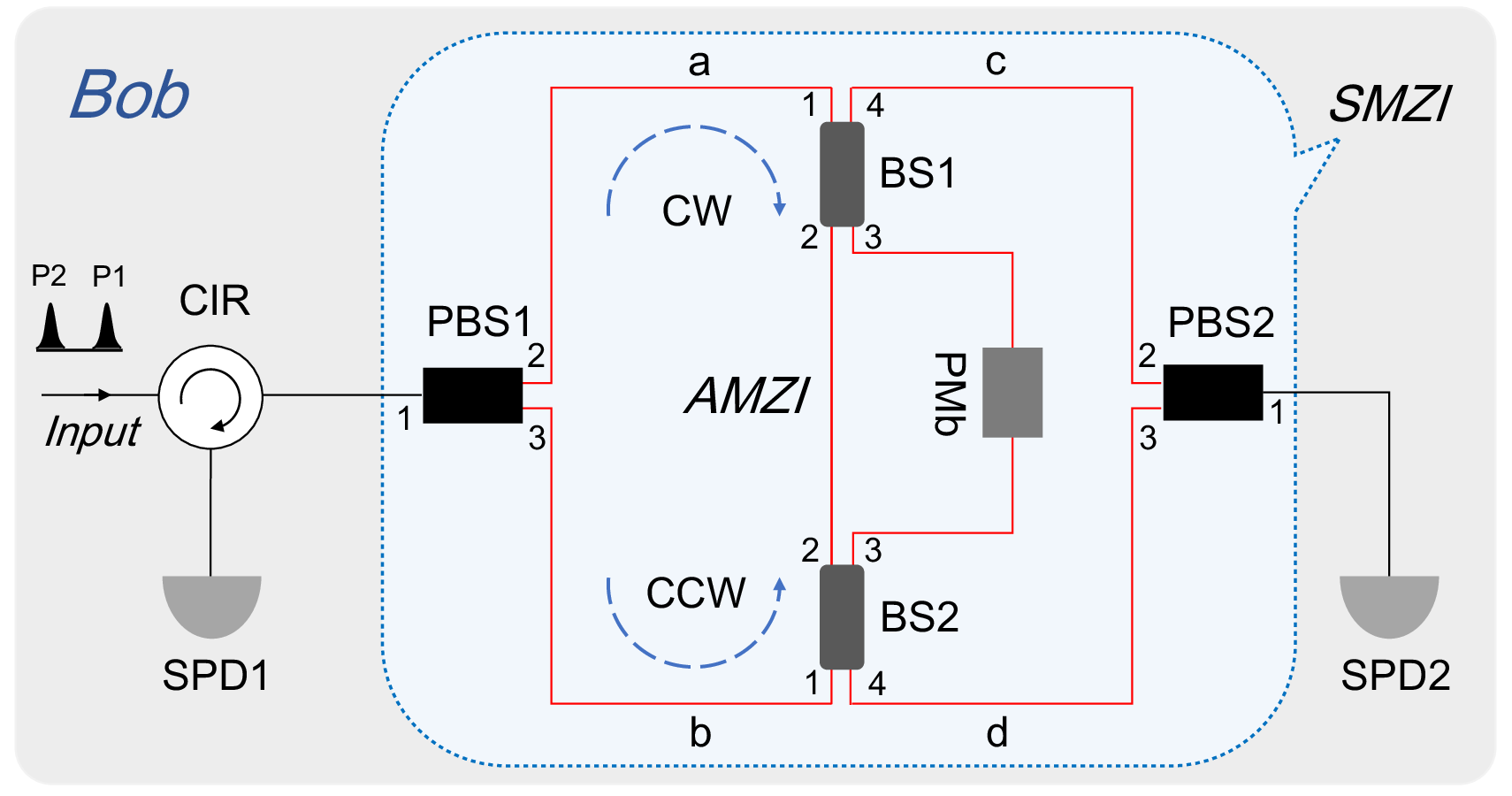}
\caption{Configuration of the receiver, the details of a SMZI is shown in the blue dashed box. The connection PMFs between AMZI and the ports of PBS1 or PBS2 are denoted by $a \sim d$, respectively. The red lines represent PMFs, and black lines represent SMFs.}
\label{SMZI}
\end{figure}

The detailed configuration of the receiver is illustrated in Fig. \ref{SMZI}. The SMZI is composed of two PBSs and an AMZI, which can be regarded as the combination of an AMZI and a Sagnac loop. After passing through a three-port circulator, the input pulses are separated to be two orthogonal polarization components by PBS1. The horizontal part ($H$) transmits PBS1 from port 1 to port 2, propagates along the clockwise (CW) direction, and then enters the AMZI. Meanwhile, the vertical part ($V$) is reflected by PBS1 into its port 3, propagates along the counterclockwise (CCW) direction, and enters the AMZI from the opposite direction. By setting the fiber length of $a\sim d$ as $l_a=l_b$ and $l_c=l_d$, and by placing PMb in the middle of the long arm of AMZI, the two parts of the input pulses would arrive at PMb simultaneously. As the output ports of common commercial polarization beam splitter are both aligned to the slow axis of PMFs, the polarization of photons would be fixed along the slow axis of PMFs in the SMZI. As a result, the two parts of the input pulses will obtain an identical phase shift when passing through PMb. The output pulses from two opposite directions of AMZI are recombined at PBS1 (PBS2) and detected by SPD1 (SPD2) simultaneously. Here the accuracy of fiber lengths should ensure that the recombined pulses cannot be time resolved with the detectors, which can be achieved easily.

We now present an analysis of the polarization insensitivity of the SMZI with the Jones matrix. As mentioned above, the polarization direction of photons from port 3 of PBS1 is aligned to the slow axis of the PMF with a rotation of $90^{\circ}$, the Jones matrix of the rotation operation is
\begin{equation}\label{R}
R = \left[ {\begin{array}{*{20}{c}}
0 & -1 \\
1 &  0
\end{array}} \right].
\end{equation}
Noted that the Jones matrices of the forward and backward propagation of the rotation operation are the same. Furthermore, the effect of the PMFs in the SMZI will be omitted in the analysis below, in view of the fact that the Jones matrix of PMF can be regarded as a unit matrix. For simplicity, we also omit the loss of all elements. The Jones matrices of the polarization-maintaining beam splitter (BS) and the PBS are given by \cite{SHSun}
\begin{equation}
\begin{split}
B_{12}^n & = B_{21}^n = \frac{1}{{\sqrt 2 }}\left[ {\begin{array}{*{20}{c}}
1&0\\
0&1
\end{array}} \right], B_{13}^n = B_{31}^n = \frac{i}{{\sqrt 2 }}\left[ {\begin{array}{*{20}{c}}
1&0\\
0&1
\end{array}} \right], \\
 P_{12}^n & = P_{21}^n = \left[ {\begin{array}{*{20}{c}}
1&0\\
0&0
\end{array}} \right], P_{13}^n =P_{31}^n = \left[ {\begin{array}{*{20}{c}}
0&0\\
0&1
\end{array}} \right],
\end{split}
\end{equation}
where $B_{jk}^n$ ($P_{jk}^n$ ) is the Jones matrix of the $n$th BS (PBS) and the subscript represents that the photons input from port $j$ and output from port $k$ (the definition of each port is given in Fig. \ref{SMZI}).

The polarization of pulses $P1$ and $P2$ emitted from Alice become unpredictable when arriving at Bob's site after transmission in the channel, its Jones vector can be written as a normalized formalization,
\begin{equation}\label{Ein}
E_{in} = \left[ {\begin{array}{*{20}{c}}
{\cos \theta }\\
{{e^{i\beta }}\sin \theta }
\end{array}} \right],
\end{equation}
in which $\theta$ is the angle to the horizontal polarization, and $\beta$ is the phase delay between the horizontal and vertical polarization.

Thus for the output port of the SMZI from PBS1, it is easy to write the transformation matrices of two interference pulses as follows:
\begin{equation}
\begin{split}
T_L &= T_L^{CW} + T_L^{CCW} \\
& = P_{31}^1 \cdot R \cdot B_{31}^2 \cdot PMb \cdot B_{13}^1 \cdot P_{12}^1 \\
 &+ P_{21}^1 \cdot B_{31}^1 \cdot PMb \cdot B_{13}^2 \cdot R \cdot P_{13}^1    \\
& =  - \frac{{{e^{i{\varphi _b}}}}}{2}\left[ {\begin{array}{*{20}{c}}
0&{ - 1}\\
1&0
\end{array}} \right],   \\
T_S &= T_S^{CW} + T_S^{CCW} \\
&= P_{31}^1 \cdot R \cdot B_{21}^2 \cdot B_{12}^1 \cdot P_{12}^1  \\
&+ P_{21}^1 \cdot B_{21}^1 \cdot B_{12}^2 \cdot R \cdot P_{13}^1   \\
&= \frac{1}{2}\left[ {\begin{array}{*{20}{c}}
0&{ - 1}\\
1&0
\end{array}} \right]
\end{split}
\end{equation}
where $T_L$ ($T_S$) is the transformation matrix for pulse $P1$ ($P2$ ) taking the path of the long (short) arm of the AMZI, and $\varphi_b$ is the phase modulated by PMb. Considering the phase difference $\varphi_a$ between $P1$ and $P2$ induced by PMa, the output of the SMZI from PBS1 can be written as
\begin{align}\label{Eout1}
E_{out1} &= \left( {{T_L} + {e^{i{\varphi _a}}}{T_S}} \right){E_{in}}  \nonumber \\
&=\frac{1}{2}\left( {{e^{i{\varphi _a}}} - {e^{i{\varphi _b}}}} \right)\left[ {\begin{array}{*{20}{c}}
{ - {e^{i\beta }}\sin \theta }\\
{\cos \theta }
\end{array}} \right]
\end{align}
Thus the interference output is expressed as
\begin{equation}\label{I1}
{I_{out1}} =E_{out1}^\dagger \cdot E_{out1}= \frac{1}{2}\left[ {1 - \cos \left( {{\varphi _a} - {\varphi _b}} \right)} \right]
\end{equation}

Following a similar procedure, we can obtain the output state of the SMZI from PBS2 as
\begin{equation}\label{Eout2}
{E_{out2}} =\frac{i}{2}\left( {{e^{i{\varphi _a}}} + {e^{i{\varphi _b}}}} \right)\left[ {\begin{array}{*{20}{c}}
{ - {e^{i\beta }}\sin \theta }\\
{\cos \theta }
\end{array}} \right]
\end{equation}
and we can obtain the corresponding interference outcome as
\begin{equation}\label{I2}
{I_{out2}} = \frac{1}{2}\left[ {1 + \cos \left( {{\varphi _a} - {\varphi _b}} \right)} \right]
\end{equation}
It is obvious from formulas (\ref{Eout1})$\sim$(\ref{I2}) that the intensity of the output states is independent of the input polarization, indicating the polarization insensitivity of the SMZI. Interestingly, the outcomes are the same as those obtained in the case in which Bob uses an AMZI with polarization calibration, as displayed in Fig. \ref{Fig:scheme} a.

In addition to the intrinsic feature of polarization insensitivity as described above, there are two more merits of the SMZI-based QKD system. First, in the SMZI, two parts of the input pulse from the opposite direction pass through PMb with the same polarization and are modulated by it simultaneously, so the only requirement for PMb is to support one polarization and bi-directional modulation, which can be satisfied by standard off-the-shelf products. It is also compatible with high speed operations by employing a simple phase modulation scheme. Second, the insertion loss is comparable with that of the traditional AMZI, since the input pulse is equivalent to passing through PMb once. As a result, we found that the performance of our QKD scheme is comparable with that of a FSMI-based one \cite{WS}.

\section{Experiment}
\begin{figure}[!ht]
\centering
\includegraphics[width=1\columnwidth]{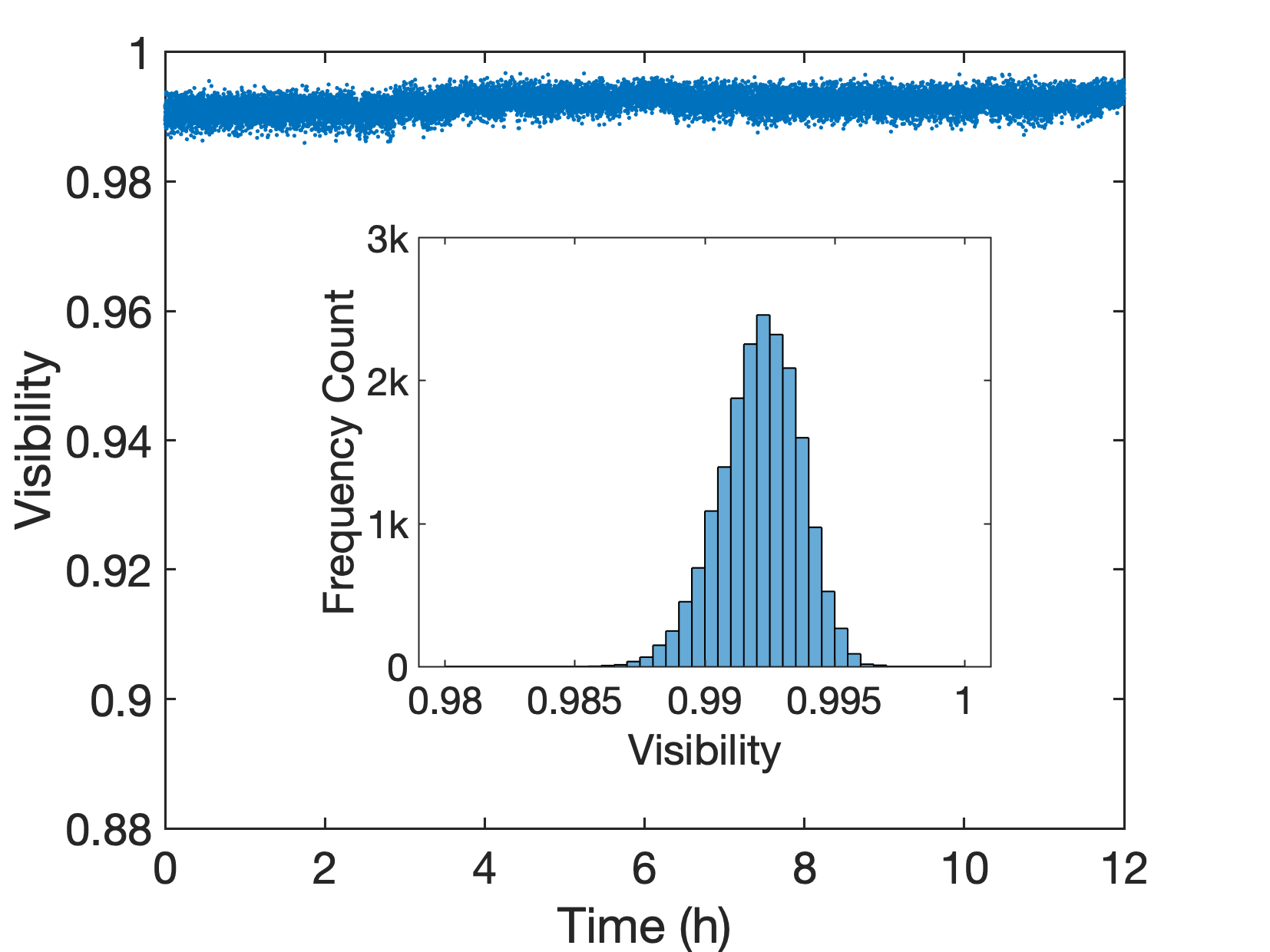}
\caption{Measured visibility of the SMZI-based QKD system over 12 h, the inset is the histogram of the visibility.}
\label{vi}
\end{figure}

To evaluate its performance, the SMZI scheme is applied into a phase-coding BB84 QKD system. First, the interference fringe visibility of the SMZI-based QKD system over a-50 km fiber channel is tested for 12 h. In order to verify its stability against polarization disturbance, a polarization scrambler module (N7788B, Keysight) is inserted along the channels to randomize the polarization with a scrambling speed of 2 rad/s. By scanning the driving voltage of Bob's PM from -5 V to 5 V with a step of 0.05 V in each round, we obtain the sinusoidal curve of SPD counts. Then, the corresponding interference fringe visibility can be calculated by
\begin{equation}\label{V}
V=\frac{ C_{max} - C_{min} }{ C_{max} + C_{min} },
\end{equation}
where $C_{max}$ and $C_{min}$ are maximum and minimum counts, respectively. Fig. \ref{vi} shows measured visibility over 12 h, as well as the histogram within the measurement time. The average visibility achieves $99.21 \pm 0.15 \%$, which demonstrates the stability of the QKD system with the presence of polarization scrambling.

The QKD system based on the SMZI scheme is implemented at a repetition frequency of 40 MHz. The pseudo-single-photon source is realized by a pulsed distributed feedback (DFB) laser, which works at 1550.12nm with a 50 ps pulse width. The IM is used to modulate the mean photon numbers of each pulse to generate the signal and decoy states, while the vacuum pulses are generated by shutting down the laser, which are required in the ``Weak+Vacuum'' decoy-state method \cite{3decoy}. The signal pulses are used for key generation, while the other two classes of pulses are used as decoy pulses to characterize the quantum channel. The double-channel SPD used in the receiver is based on InGaAs/InP avalanche photodiodes and is operated in the gated Geiger mode, the average detection efficiency of the two channels is $10\%$, the average after-pulse probability is $0.5\%$, and the total dark count rate is $3.5\times10^{-6}$ per gate. The insertion loss of the receiver (including the loss of noninterference parts) is measured to be 5.95 dB. To carry out the standard BB84 protocol, we perform the postprocessing in software. Error correction is carried out by using the cascade algorithm (including error verification) with an efficiency $f_{EC}$ of about 1.14, and privacy amplification is carried out by using Toeplitz matrix multiplication. Taking the finite-key effect into account, the block size of privacy amplification is set to be $2^{20}$, and the secure key rate (SKR) is calculated based on the security analysis in Ref. \cite{Lim} with a failure probability of $10^{-10}$. The intensity of signal (decoy) pulses and the corresponding probability are listed in Table \ref{table1}.

\begin{table}[hbp]
  \centering
  \caption{Intensities of signal pulses and decoy pulses, and their corresponding probabilities. $\mu$, $\nu_1$, and $\nu_2$ are the mean photon numbers of the signal, decoy, and vacuum pulses, and $p_{\mu}$, $p_{\nu_1}$, and $p_{\nu_2}$ are the corresponding probabilities, respectively. }
\begin{tabular}{cccccccc}  \hline  \hline
     $\mu$  & $\nu_1$ & $\nu_2$ & $p_{\mu}$ & $p_{\nu_1}$  & $p_{\nu_2}$ \\ \hline
      0.6 & 0.1 & 0 & 29/32 & 2/32 & 1/32 \\ \hline  \hline
  \end{tabular}
 \label{table1}
\end{table}

To guarantee the stability of the QKD system over a long-transmission fiber link for long-term operation, two realistic issues need to be dealt with: (i) channel transmission instability, and (ii) phase drift. The first one, channel transmission instability, mainly comes from polarization characteristics' variation and fiber length drift. The former adds to the quantum bit error rate (QBER) and can be avoided by employing the SMZI-based scheme. The latter leads to the photon arrival time at the detectors moving outside of the active time window, causing a drop in the bit rate, which can be compensated for by using the detector count rates as a feedback signal to adjust the delay position of the detector gate. As for phase drift, a real-time phase-tracking method \cite{Phacom} is applied in our QKD system to ensure long-term operation. In the phase-tracking method, only mismatched-basis data are used to calculate the phase drift parameter and adjust the driving voltages of PM, which can be performed after the basis-sifting stage. The announcement of mismatched-basis data and the acquisition of the phase parameter cost rather limited time. In consequence, the system can operate continuously without being interrupted by the phase compensation process.

\section{Results and discussion}
\begin{figure}[!ht]
\centering
\includegraphics[width=1\columnwidth]{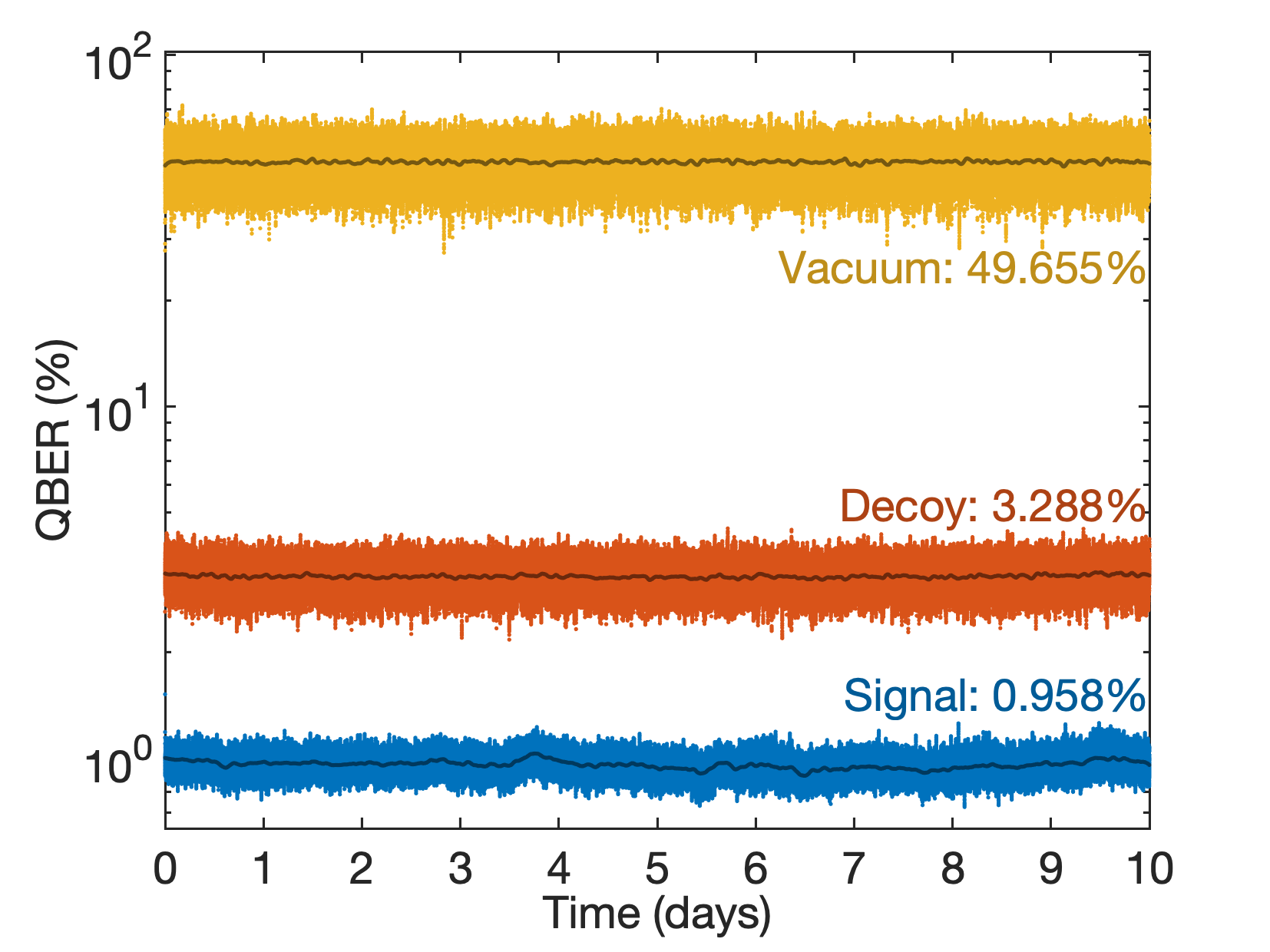}
\caption{QBERs of the signal, decoy and vacuum pulses over 12.6-dB channel loss.}
\label{qber}
\end{figure}

We implement the QKD system over a 50-km fiber channel with continuous polarization scrambling (2 rad/s) for long-term running test. The insertion loss of the 50-km fiber spool and the polarization scrambler module are 9.2 dB and 3.4 dB, respectively, resulting in a 12.6-dB total channel loss. QBERs of the signal, decoy and vacuum states over 10 days of continuous operation are shown in Fig. \ref{qber}, the solid lines are average values of data for every 3 h. The average QBERs for the signal, decoy, and vacuum pulses are $0.958\%$, $3.288\%$ and $49.655\%$, respectively. Fig. \ref{qber} clearly shows that the QBER for the signal pulses stays rather low and stable for 10 days, validating the effectiveness of our system, and the results agree well with the stability of interference fringe visibility.

\begin{figure}[!ht]
\centering
\includegraphics[width=1\columnwidth]{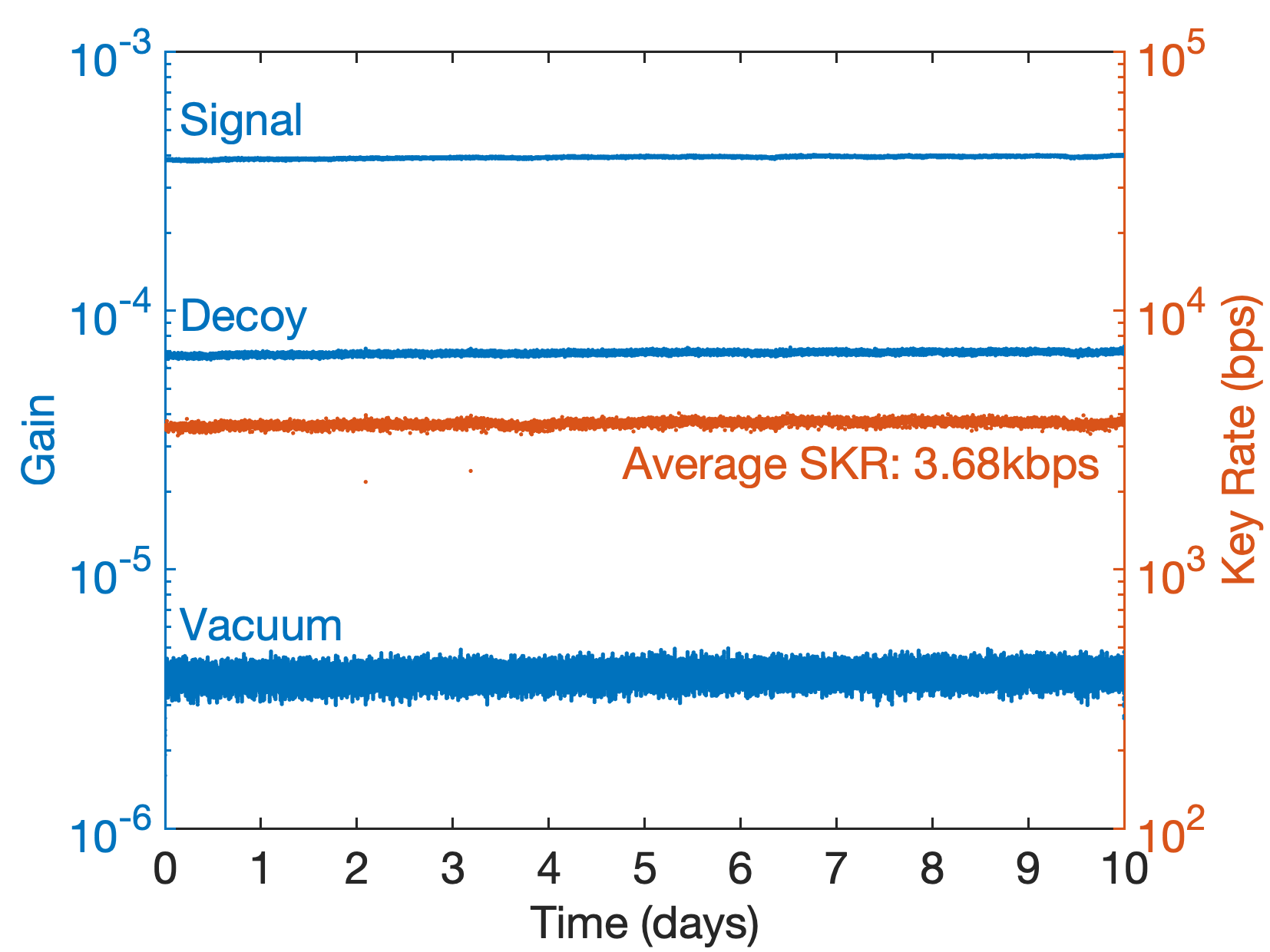}
\caption{Overall gains (left axis) of the signal, decoy and vacuum pulses, and the secure key rate (right axis) over 12.6 dB channel loss.}
\label{gain}
\end{figure}

Fig. \ref{gain} displays the corresponding overall gains (detection probability per pulse sent) of three states and the obtained SKR. As is shown, the average SKR is 3.68 kbps and remains relatively stable for 10 days. The results clearly show the long-term stability of our system under channel disturbance, indicating a comparable performance of the FSMI-based QKD system \cite{WS} in terms of robustness.

\begin{table}[hbp]
  \centering
  \caption{Experimental results over different channel losses.}
\begin{tabular}{cccccc}  \hline  \hline
     Channel loss/dB        &      10      &    12.6     &    15       &       20    &    25     \\ \hline
     Sifted key rate/bps   &  21969.0  &  11804.8  &   7299.7  &   2408.6  &   838.8   \\
     Secure key rate/bps  &  6894.9   &  3675.9    &   2128.3  &  537.0     &  54.6      \\
     QBER/$\%$              &  0.899     &   0.958     &  1.181     &  1.991     &  4.205    \\ \hline  \hline
  \end{tabular}
 \label{table2}
\end{table}

The performance of our QKD system over different channel losses is further investigated. By setting the channel loss to be 10, 15, 20 , and 25dB, respectively, the corresponding sifted key rates and QBERs for signal states, as well as the average SKR are obtained. The experimental and simulated results are shown in Fig. \ref{skr} and Table \ref{table2}. The gains and error rates model used in the simulation can be found in the Appendix. Remarkably, the experimental results are in good agreement with simulations.

\begin{figure}[!ht]
\centering
\includegraphics[width=1\columnwidth]{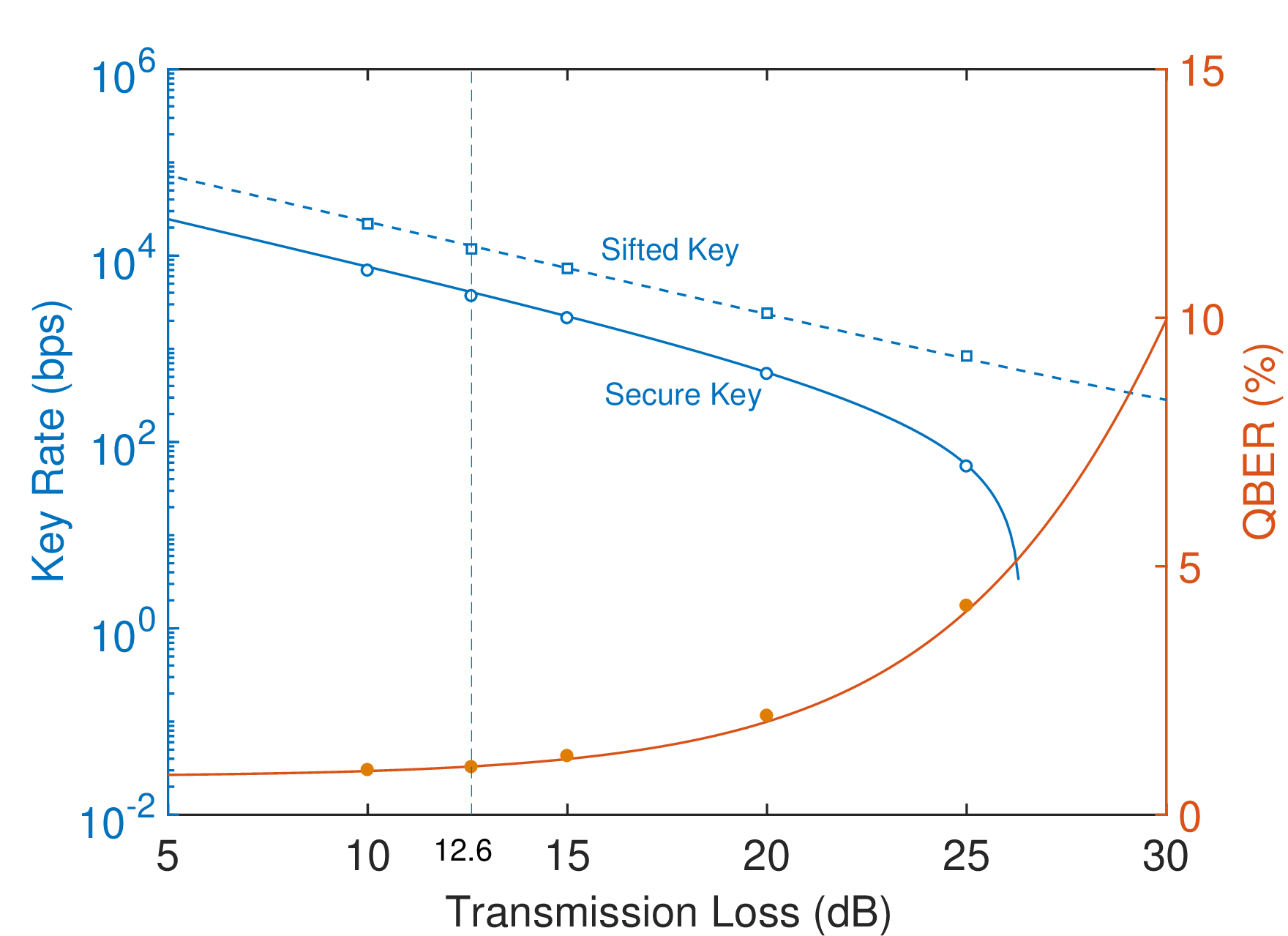}
\caption{Experimental and simulated results over different channel losses. Circles, squares, and dots are, respectively, the experimental results of SKRs, sifted key rates, and QBERs for signal states. Solid, dashed and dotted line individually denotes the simulated results of SKRs, sifted key rates, and QBERs for signal states.}
\label{skr}
\end{figure}

\section{Conclusion}
In summary, we have proposed a polarization-insensitive SMZI-based QKD scheme and theoretically analyzed its stability against polarization disturbance. We developed a QKD system based on SMZIs and demonstrated its performance for 10 days over a 50-km fiber channel with polarization scrambling. Experimental results show that this system is able to operate continuously and autonomously over long periods of time and remain a stable and low level of QBER against channel disturbance. Moreover, the maximum transmission loss of the system can exceed 25 dB, signifying the equivalent maximum transmission distance can be achieved as more than 125 km. We believe that our SMZI-based QKD system is suitable for practical deployment in field environments.

\section*{APPENDIX}
In this Appendix, we present the gains and error rates model in the simulation.

The gain (detection rate) without after-pulse contributions for intensity $k$ is given by
\begin{equation}\label{Qk0}
Q_k^0=1-(1-P_{dc})e^{-k\eta}, (k=\mu,\nu_1,\nu_2)
\end{equation}
where $\eta$ is the overall transmission and detection efficiency between Alice and Bob, i.e., $\eta=10^{-0.1(L_C + L_B)}\eta_d$, where $L_B$($L_C$) is the loss of the receiver (fiber channel).

Then the gain including after-pulse contributions for intensity $k$ is modified to be
\begin{equation}\label{Qk}
Q_k=Q_tP_{ap} + Q_k^0,
\end{equation}
where $Q_t=\sum_k p_kQ_k^0$ is the average gain for a single pulse sent by Alice, and $p_k$ is the probability that Alice selects the intensity $k$.

The corresponding error rate for intensity $k$ is expressed as
\begin{equation}\label{Ek}
 E_k=\frac{\frac{1}{2}P_{dc}+e_{mis}(1-e^{-k\eta})+\frac{1}{2}Q_tP_{ap}}{Q_k},
\end{equation}
where $e_{mis}$ is the optical misalignment error rate.

\end{document}